\documentclass[aps,prd,floatfix,twocolumn,superscriptaddress,a4paper]{revtex4-1}

\pdfoutput=1

\usepackage{amsmath}
\usepackage{amssymb}
\usepackage{graphicx}
\usepackage{hyperref}
\usepackage{color}
\usepackage{xcolor}
\usepackage[T1]{fontenc} 
\usepackage{slashed}
\usepackage[utf8]{inputenc}

\usepackage[normalem]{ulem} 

\definecolor{blus}{cmyk}{1,1,0,0.6}
\hypersetup{colorlinks,bookmarksopen,bookmarksnumbered,linkcolor=blus,pdfstartview=FitH,urlcolor=blus,citecolor=blus}

\def\hc{\mathrm{h.c.}}

\allowdisplaybreaks

\newcommand{\AddrIFIC}{%
Instituto de F\'{\i}sica Corpuscular (CSIC-Universitat de Val\`{e}ncia),
Apdo. 22085, E-46071 Valencia, Spain
}

\newcommand{\AddrZur}{%
Physik-Institut, Universit{\"a}t Z{\"u}rich, CH-8057 Z{\"u}rich, Switzerland
}

\preprint{IFIC/19-31, ZU-TH-36/19}
\begin{document}

\title{
Strong CP problem with low-energy emergent QCD: The 4321 case
}

\author{Javier Fuentes-Mart\'in}\email{fuentes@physik.uzh.ch}
\affiliation{\AddrZur}

\author{Mario Reig} \email{mario.reig@ific.uv.es}
\affiliation{\AddrIFIC}

\author{Avelino Vicente} \email{avelino.vicente@ific.uv.es}
\affiliation{\AddrIFIC}

\begin{abstract}
We analyze the strong CP problem and the implications for axion physics in the context of $U_1$ vector leptoquark models, recently put forward as an elegant solution to the hints of lepton flavor universality violation in $B$-meson decays. It is shown that in minimal gauge models containing the $U_1$ as a gauge boson, the Peccei-Quinn solution of the strong CP problem requires the introduction of two axions. Characteristic predictions for the associated axions can be deduced from the model parameter space hinted by $B$-physics, allowing the new axion sector to account for the dark matter of the Universe. We also provide a specific ultraviolet completion of the axion sector that connects the Peccei-Quinn mechanism to the generation of neutrino masses.
\end{abstract}

\maketitle

\section{Introduction}

The recent indications of Lepton Flavor Universality Violation (LFUV) in semileptonic $b\to c\tau\nu$~\cite{Lees:2013uzd,Aaij:2015yra,Hirose:2016wfn,Aaij:2017deq,Abdesselam:2019dgh} and $b\to s\ell\ell$~\cite{Aaij:2014ora,Aaij:2017vbb,Aaij:2019wad,Abdesselam:2019wac} transitions are at present one of the most interesting hints of New Physics (NP). Even though no individual measurement presents a high statistical significance, the global picture is very compelling: the internal consistency of the data is remarkable~\cite{Alguero:2019ptt,Aebischer:2019mlg,Ciuchini:2019usw,Datta:2019zca,Shi:2019gxi,Murgui:2019czp} and, once combined, the significance of the LFUV observables exceeds $3.7\sigma$ in $b\to s\ell\ell$ and $3.1\sigma$ in $b\to c\ell\nu$. A common origin for these deviations is not obvious, but it is very appealing from the theoretical point of view. If confirmed as clear signals of physics beyond the Standard Model (SM), they would point to nontrivial dynamics at the TeV scale, possibly linked to a solution of the SM flavor puzzle~\cite{Barbieri:2015yvd,Buttazzo:2017ixm,Barbieri:2019zdz,Bordone:2017bld,Greljo:2018tuh,Bordone:2018nbg,Cornella:2019hct}.

The $U_1$ vector leptoquark, transforming as $(\mathbf{3},\mathbf{1},2/3)$ under the SM gauge group, has revealed itself as an excellent candidate for this task. The search for a renormalizable model that contains the $U_1$ has led to
the so-called \textit{4321 models}~\cite{Georgi:2016xhm,Diaz:2017lit,DiLuzio:2017vat,Blanke:2018sro,DiLuzio:2018zxy,Bordone:2017bld,Greljo:2018tuh,Bordone:2018nbg,Cornella:2019hct}, where the SM gauge group is extended to $\mathcal{G}_{4321}\equiv SU(4)\times SU(3)^\prime\times SU(2)_L\times U(1)^\prime$. This is the smallest gauge group allowing for a TeV-scale $U_1$ as a gauge boson while remaining consistent with the stringent bounds from high-$p_T$ data, see~\cite{Baker:2019sli} for a recent discussion. In $4321$ models, the SM color group is embedded as $SU(3)_c = \left[SU(3)_4\times SU(3)^\prime\right]_{\rm diag}$, where $SU(3)_4$ is a subgroup of $SU(4)$.  An interesting feature of 4321 models is that QCD arises as a low-energy interaction emerging from the product of two non-Abelian groups. Therefore, it is tempting to study why CP is conserved in the strong sector of these theories.

In this article we show how a generalization of the Peccei-Quinn (PQ) mechanism is applied to the 4321 models to solve the strong CP problem. A characteristic feature of this mechanism is that two axions are predicted. Many properties of these axions are deduced from the parameter space required to accommodate the deviations in $B$-meson decays. In particular, we find that one of the axions remains QCD-like, while the other is much heavier. In the last part we provide an ultraviolet completion for the proposed PQ mechanism that connects this solution to the generation of neutrino masses.

\section{The Strong CP problem in 4321 models}
We are interested only in the sector of the theory involving QCD interactions: $SU(4)\times SU(3)^\prime$. We denote the corresponding gauge bosons respectively as $H_\mu^A$ and $C_\mu^a$ with indices $A=1,\dots,15$ and $a=1,\dots,8$, and the gauge couplings as $g_4$ and $g_3$. The spontaneous symmetry breaking (SSB) $SU(4)\times SU(3)^\prime\to SU(3)_c$, is assumed to take place at a scale $M\sim$~TeV.  After SSB, the gauge boson mass eigenstates are given in terms of the original gauge bosons by
\begin{align}
\begin{aligned}
G^a_\mu&=\cos\gamma\, C_\mu^a + \sin\gamma\, H_\mu^a \,,\\
G^{\prime\,a}_\mu&= -\sin\gamma\, C_\mu^a + \cos\gamma\, H_\mu^a\,,
\end{aligned}
\end{align}
with $\tan\gamma=g_3/g_4$. Here $G$ corresponds to the QCD gluons and $G^\prime$ is a massive color-octec vector, which we denote as coloron. For the QCD coupling we have 
\begin{align}\label{eq:gs_matching}
g_s=\frac{g_3\,g_4}{\sqrt{g_3^2+g_4^2}}\,.
\end{align}
It is important to note that $g_{3,4}>g_s$, and that any of these couplings can be significantly larger than $g_s$ at the scale $M$. This is important when computing nonperturbative contributions to the the axion potential. 

The relevant Lagrangian for the QCD $\theta$-term is
\begin{equation}
\mathcal{L}=\frac{\theta_4\alpha_4}{8\pi}H^A_{\mu\nu}\tilde{H}^{A\,\,\mu\nu}+\frac{\theta_3\alpha_3}{8\pi}C^a_{\mu\nu}\tilde{C}^{a\,\,\mu\nu}\,,
\end{equation}
which for the QCD gluons gives
\begin{equation}\label{eq:thetaQCD}
\mathcal{L}\supset(\theta_4+\theta_3)\,\frac{\alpha_s}{8\pi}\,G^a_{\mu\nu}\tilde{G}^{a\,\,\mu\nu}\,,
\end{equation} 
and where $\alpha_i=g_i^2/4\pi$. From this Lagrangian it is clear that the tree-level value of the QCD $\theta$-term is $\bar\theta_{\rm QCD}=\bar\theta_4+\bar\theta_3$. Here $\bar \theta_i$ is defined as usual as the effective $\theta_i$ angle in the basis in which the fermion masses are real. As in the SM, loop-level corrections to this angle proportional to the CKM phase are negligible~\cite{Ellis:1978hq,Dugan:1984qf}. Additional $\theta$-terms involving the coloron also mediate CP-violating effects. Even though these are suppressed by the coloron mass, they would lead to an unacceptably large neutron electric dipole moment if the $\theta_i$ angles are of $\mathcal{O}(1)$.

QCD contributions to the axion potential are the same as in the SM. Thus, it might seem that the strong CP problem could still be solved with a single axion. This is not the case for two reasons. First, due to the potentially large coloron-induced contributions. And second, because of the new short-distance nonperturbative effects induced by the $SU(4)$ and $SU(3)^\prime$ factors, see~\cite{Agrawal:2017ksf,Agrawal:2017evu,Gaillard:2018xgk} for a recent discussion and~\cite{Holdom:1982ex,Flynn:1987rs} for a similar discussion in a different context. These new effects generate shift-symmetry-breaking potential terms that destabilize the single-axion solution to the strong CP problem.  We compute them in the next section. A straightforward solution consists in including two axions, each relaxing one of the $\theta_i$ angles. The two-axion potential is given by
\begin{align}\label{eq:axPot}
\begin{aligned}
V_{\rm axion}&\approx-\Lambda_{\rm QCD}^4\,\cos\left(\frac{a_1}{f_{a_1}}+\frac{a_2}{f_{a_2}}-\bar\theta_{\rm QCD}\right)\\
&\quad-\Lambda_{SU(3)}^4\,\cos\left(\frac{a_1}{f_{a_1}}-\bar\theta_3\right)\\
&\quad-\Lambda_{SU(4)}^4\,\cos\left(\frac{a_2}{f_{a_2}}-\bar\theta_4\right)\,,
\end{aligned}
\end{align}
where the effective scale $\Lambda_{\rm QCD}$ encodes the QCD contributions to the axion potential which is estimated using $\chi$PT techniques~\cite{Weinberg:1977ma,Wilczek:1977pj} to be $\Lambda_{\rm QCD}\approx 77$~MeV. The minimum of this potential,
\begin{align}
\left\langle\frac{a_1}{f_{a_1}} -\bar\theta_3\right\rangle=0\,,\qquad\qquad
\left\langle\frac{a_2}{f_{a_2}} -\bar\theta_4\right\rangle=0\,,
\end{align}
is CP conserving. This solution is guaranteed to solve the strong CP problem by a generalization of the Vafa-Witten theorem~\cite{Vafa:1984xg}, and also removes coloron-mediated CP-violating contributions via the $\theta$-terms. The need for a second axion therefore constitutes a smoking-gun signature of this class of models. 

The properties of these axions are determined by the $SU(4)$ and $SU(3)^\prime$ nonperturbative effects parametrized by $\Lambda_{SU(4)}$ and $\Lambda_{SU(3)}$, which are fixed once the matter content and the gauge couplings at the TeV scale are specified. As we show later, in the models we are interested in, $\Lambda_{SU(3)}\approx0$ and $\Lambda_{SU(4)}\gg\Lambda_{\rm QCD}$. In this limit and assuming no large hierarchies between the axion decay constants $f_{a_1}$ and $f_{a_2}$, we obtain the following physical axion states
\begin{align}
a_\ell\approx a_1-\epsilon\,a_2\,,\qquad a_h\approx a_2+\epsilon\,a_1\,,
\end{align}
where $\epsilon=f_{a_2}/f_{a_1}\times(\Lambda_{\rm QCD}/\Lambda_{SU(4)})^4$. The corresponding axion masses in the same limit, with $f_\ell\approx f_{a_1}$ and $f_h\approx f_{a_2}$, read
\begin{align}\label{mass_axion}
m_{a_\ell}\approx\frac{\Lambda_{\rm QCD}^2}{f_\ell}\,,\qquad m_{a_h}\approx\frac{\Lambda_{SU(4)}^2}{f_h}\,,
\end{align}
so one of the axions remains QCD-like, while the other one receives a much larger mass proportional to the nonperturbative scale $\Lambda_{SU(4)}$. Due to the structure of the axion potential, the two axions present the same axion-pion mixing, which coincides with that of the QCD axion. Moreover, since the $\theta_i$ angles of both groups contribute to $\theta_{\rm QCD}$ with the same weight, both axions are predicted to have the same coupling to $G\tilde G$, again coinciding with that of the QCD axion, and hence also to nucleons in the absence of axion couplings to fermions.

\section{Small-size instantons}
Contributions from $SU(4)$ and $SU(3)^\prime$ nonperturbative effects at the scale $M\sim$~TeV are potentially large. This is particularly the case when the coupling constant of one of these groups becomes much larger than the QCD coupling. We estimate the  small-size instanton (SSI) contributions to the axion potential using the dilute instanton gas approximation (DIGA)~\cite{Callan:1977gz} 
\begin{align}\label{eq:Lambda}
\Lambda_i^4\approx2\,\int_0^\infty \frac{d\rho}{\rho^5}\,\prod_{k}^{N_f}\left(\rho m_k\right)\,D[\alpha_i(1/\rho)]\,,
\end{align}
where $N_f$ is the number of fermion zero modes and $m_k$ their corresponding masses. Since the gauge groups we are interested in undergo SSB, the instanton density is described by the constrained instanton formalism~\cite{tHooft:1976snw,Affleck:1980mp}
\begin{align}
D[\alpha(1/\rho)]=C_{inst}\,\left(\frac{2\pi}{\alpha}\right)^{2N}\,e^{-2\pi/\alpha(1/\rho)-4\pi^2\rho^2\,\mathcal{V}^2}\,,
\end{align}
with
\begin{align}
C_{inst}=\frac{4}{\pi^2}\frac{2^{-2N}e^{-[1+N_s^A]\,c(1)+[N_f-N_s^F-2(N-2)]\,c(1/2)}}{(N-2)!(N-1)!}\,,
\end{align}
where $c(1)\approx0.443$, $c(1/2)\approx0.146$, $N_s^F (N_s^A)$ is the number of fundamental (adjoint) scalars coupled to $SU(N)$, with $N=3,4$, and $\mathcal{V}^2=\sum_i\,q_i\langle\Phi_i\rangle^2$, with $q_i=1/2\,(1)$ for fundamental (adjoint) scalars. The factor proportional to the vacuum expectation values (VEV) yields an exponential suppression for large instanton sizes (i.e. when $2\pi\rho\gtrsim\mathcal{V}^{-1}$), acting as an infrared regulator. Constrained instantons are therefore much better behaved than the ones in QCD, making the results of the DIGA more reliable. Finally, note that the coupling appearing in the exponential is renormalized, but the one in the pre-exponential factor is not. It is expected that at two loops the pre-exponential factor gets renormalized~\cite{Diakonov:1983hh} and thus one replaces the bare coupling by the one-loop running coupling, and the one-loop coupling in the exponential by the two-loop coupling. Taking this into account, the final expression reads
\begin{widetext}
\begin{align}\label{eq:L4}
\begin{aligned}
\Lambda_i^4&\approx C_{inst}\,\left(\frac{2\pi}{\alpha_i(M)}\right)^{2N}\,e^{-2\pi/\alpha_i(M)}\int_0^\infty \frac{d\rho}{\rho^5}\,\prod_{k}^{N_f}\left(\rho\, m_k\right)(\rho M)^{b_i}\,\left(1-b_i\,\frac{\alpha_i(M)}{2\pi}\ln(\rho M)\right)^{2N-b_i^\prime/b_i}\,e^{-4\pi^2\rho^2\mathcal{V}^2}\,,
\end{aligned}
\end{align}
\end{widetext}
where $b_i=11/3\,N+\dots$ and $b_i^\prime=17/3\,N^2+\dots$ correspond to the one- and two-loop coefficients of the $SU(N)$ beta functions which can be found in~\cite{Machacek:1983tz,Jones:1981we}. 

\begin{figure}
\centering
\includegraphics[width=0.48\textwidth]{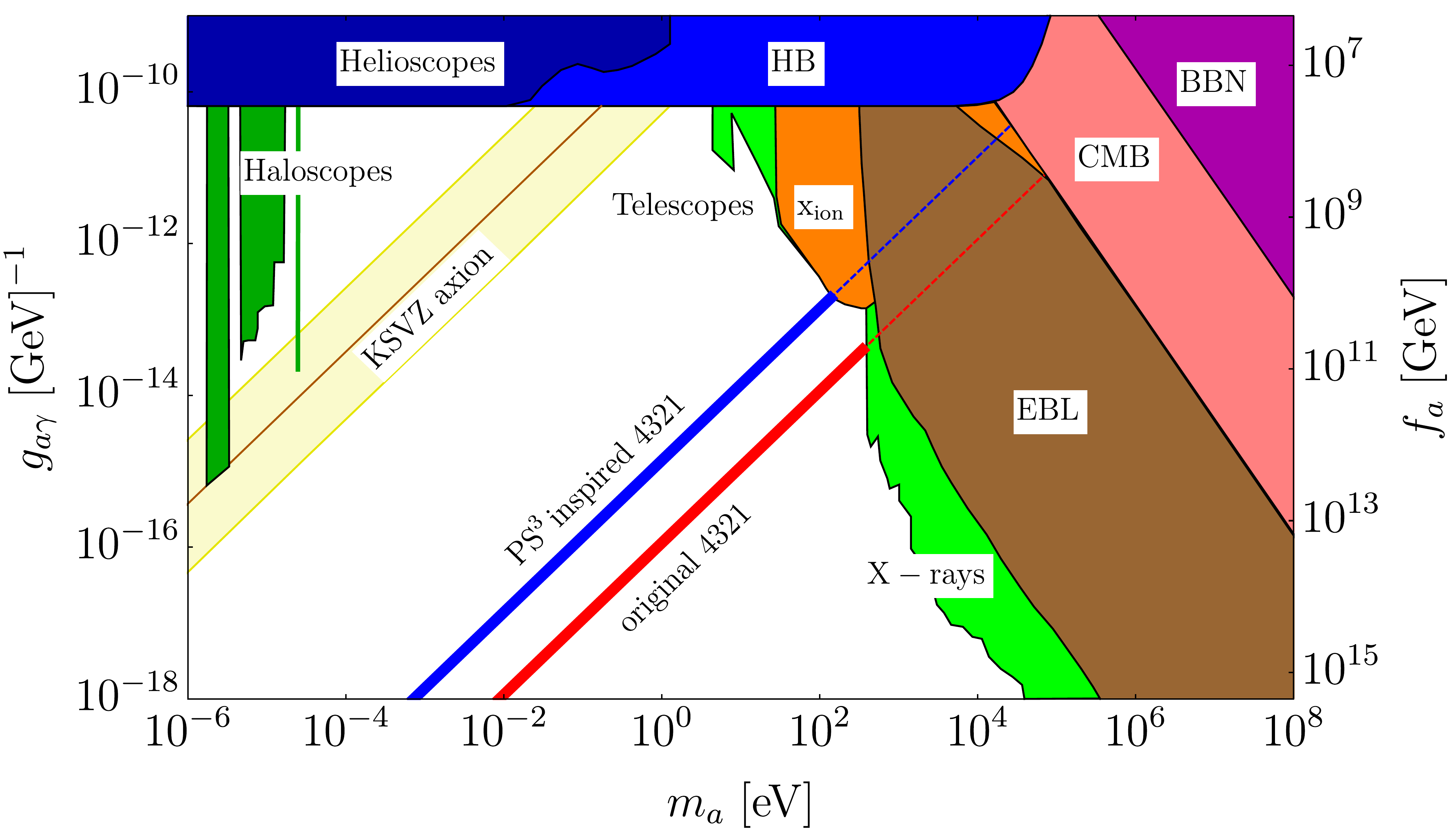}
\caption{Axion constraints in the ($g_{a\gamma}$,$m_a$) plane.}\label{fig:axionBounds}
\end{figure}

\section{Two 4321 examples from $B$-physics}
We now evaluate the expression above in 4321 models aimed to explain the $B$-physics hints of LFUV. For concreteness, we focus on the 4321 implementations in~\cite{Cornella:2019hct,DiLuzio:2018zxy}, which we denote as \textit{$\textrm{PS}^{\,3}$ inspired 4321} and \textit{original 4321} models, respectively. These two models are characterized for having large values for the $SU(4)$ coupling, a condition that is required to avoid the strong high-$p_T$ bounds on the associated $Z^\prime$ and color-octect vectors. This in turn translates into large SSI effects in the axion potential that can easily surpass the contributions from QCD interactions. 

In the original $4321$, three vector-like families are charged under $SU(4)$, while the SM-like families are charged under $SU(3)^\prime$. On the other hand, in the $\textrm{PS}^{\,3}$ inspired $4321$, third-generation SM-like fermions are charged under $SU(4)$ together with two vector-like families, while the other two SM-like families are charged under $SU(3)^\prime$. The two solutions present distinct SSI contributions to the axion potential due to the different chiral suppressions arising from the different fermion embeddings. In particular in the $\mathrm{PS}^3$ inspired 4321 model the SSI receive an additional suppression proportional to the third-family SM fermion masses. Using the benchmark points provided in~\cite{DiLuzio:2018zxy,Cornella:2019hct}, we find
\begin{align}
\begin{aligned}
\Lambda_{SU(4)}&\approx40~\mathrm{GeV}\qquad(\mathrm{PS}^3\;\mathrm{inspired}\;4321)\,,\\
\Lambda_{SU(4)}&\approx140~\mathrm{GeV}\qquad(\mathrm{original}\;4321)\,,
\end{aligned}
\end{align}
while $\Lambda_{SU(3)}\approx0$ in both cases. This is to be compared with the scale generated by QCD interactions: $\Lambda_{\rm QCD}\approx77$~MeV. In Figure~\ref{fig:axionBounds} we show the predicted heavy axion lines in the ($g_{a\gamma}$,$m_a$) plane for the two models, assuming that the axion-photon coupling, $g_{a\gamma}$, is given exclusively by the axion-pion mixing. The QCD axion band and different astrophysical constraints, from~\cite{Irastorza:2018dyq}, are also shown. As can be seen, the model receives complementary constraints from astrophysical searches, such as monochromatic gamma lines. A heavy axion in the X-ray, Extragalactic Background Light (EBL) or $x_{\rm ion}$ regions can only give a small fraction of the Dark Matter (DM) of the Universe. This is illustrated in Figure~\ref{fig:axionBounds} with the dashed lines.

\begin{figure*}
\centering
\includegraphics[width=\textwidth]{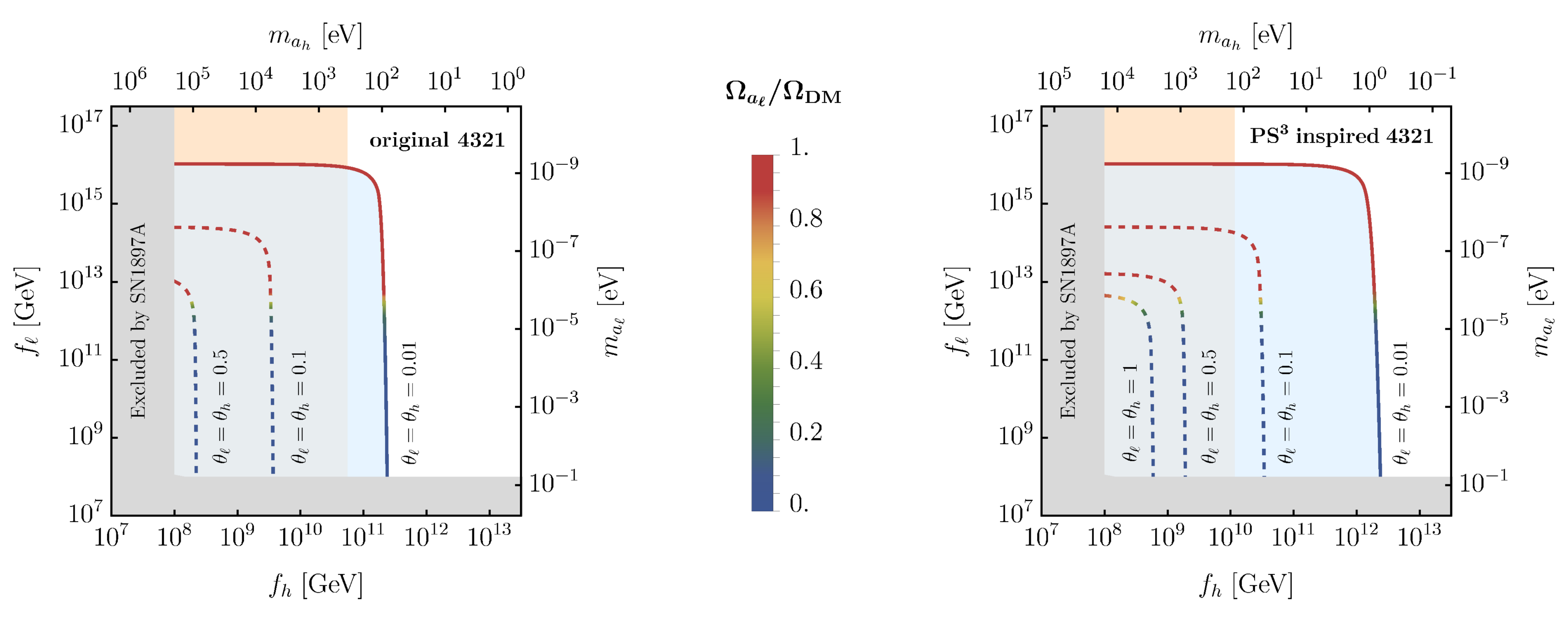}
\caption{Axion DM landscape. The blue region corresponds to values where the axions can account for 100\% of DM. The grey region is excluded from SN1987A~\cite{Raffelt:2006cw} and the orange region excludes the parameter choices of heavy axion DM dominance.}\label{fig:DMaxion}
\end{figure*}

Abundant DM in the form of cold axions can be produced via the misalignment mechanism~\cite{Preskill:1982cy,Abbott:1982af,Dine:1982ah}. In the early history of the Universe, when the temperatures are high, the axion potential is almost flat and the axion fields can take any value. As the temperature decreases, the axions start feeling the potential and begin to oscillate around its minimum, dissipating energy in the form of cold axions. The amplitude of the oscillations and hence the abundance of cold axions depends on the distance between the initial value of the axion field and the minimum of the potential, quantity that is commonly referred to as misalignment angle. This mechanism is completely determined by the shape of the axion potential at finite temperature, the zero-temperature axion masses and the initial misalignment angles, which we denote as $\theta_{\rm ini}^{\ell,h}$ for the light and heavy axions. Finite-temperature effects for the QCD-induced potential have been studied in~\cite{Wantz:2009mi,Wantz:2009it} using the instanton liquid approximation, while finite-temperature corrections to $\Lambda_{SU(4)}$ within the DIGA can be found in~\cite{Gross:1980br}. We stress that, thanks to the exponential cutoff introduced by the constrained instanton density, the DIGA is expected to yield more accurate results for $\Lambda_{SU(4)}$ than in QCD. The misalignment angles are fixed to their average values $\theta_{\rm ini}^{\ell,h}=\langle\theta\rangle\approx2.1$~\cite{diCortona:2015ldu} in the post-inflationary case, but they are free parameters in the pre-inflationary scenario. The axion DM relic density produced by this mechanism is proportional to the square of the misalignment angles and therefore it tends to zero in the limit $\theta_{\rm ini}^{\ell,h}\to0$. This is a fine-tuned configuration and it is customary to take $\theta_{\rm ini}^{\ell,h}\sim\mathcal{O}(1)$.

In Figure~\ref{fig:DMaxion}, we show the axion DM contributions for different values of the axion decay constants. The blue region corresponds to values that reproduce $(\Omega_{a_h}+\Omega_{a_l})h^2=\Omega_{DM}h^2=0.12$ with $\theta_{\rm ini}^{\ell,h}\in[2\pi,0.01]$. We also show different contours corresponding to various initial misalignment angles. An important conclusion of our analysis is that the post-inflationary scenario is strongly disfavored by data. In this scenario, the heavy axion yields either an overproduction of DM or a conflict with the astrophysical constraints. This justifies a posteriori having ignored other axion DM components, such as those from strings and/or domain walls, which are only relevant in the post-inflationary case.

\section{A possible axion ultraviolet completion}
We consider an extension of the model in~\cite{DiLuzio:2017vat,DiLuzio:2018zxy} where the generalized PQ mechanism is realized. Guided by the principle of minimality, we consider a KSVZ-like scenario \cite{Kim:1979if,Shifman:1979if,Zhitnitsky:1980tq} where the exotic PQ fermions are also responsible for the neutrino mass generation, therefore connecting the smallness of neutrino masses to the PQ scale(s)~\cite{Chikashige:1980ui,Gelmini:1980re,Kim:1981jw,Langacker:1986rj}. 

The extended particle content and global symmetries are summarized in Table~\ref{tab:bmlnur}.  The 4321 gauge symmetry is supplemented with a $U(1)_{B^\prime-L^\prime} \times U(1)_\chi$ global symmetry that acts as PQ symmetry. The first of these $U(1)$ factors can be identified with the difference between the $B^\prime$ and $L^\prime$ symmetries in~\cite{DiLuzio:2017vat,DiLuzio:2018zxy}, corresponding to the difference between baryon and lepton numbers for the SM-like fields, whereas $U(1)_\chi$ is a symmetry under which only $\chi$ and $\sigma_1$ are charged. The PQ sector of the model consists of the chiral fermions $\chi \sim (\mathbf{15},\mathbf{1},\mathbf{1},0)$ and $F \sim (\mathbf{1},\mathbf{8},\mathbf{1},0)$, and two copies of the scalar fields $\sigma_{1,2} \sim (\mathbf{1},\mathbf{1},\mathbf{1},0)$, which acquire the non-zero VEVs $\langle \sigma_i \rangle = v_{{\rm PQ}_i}$ and are responsible for the spontaneous breaking of the PQ symmetries. As in the KSVZ model, in this model the axions do not couple to SM fermions at tree level and the anomaly coefficient ratio is predicted to be $E/N=0$. This fixes the axion coupling to photons to the one given by its mixing with the pion.

\begin{table}[t]
	\centering
	\begin{tabular}{|c||cccc||cc|}
		\hline
		Field & $SU(4)$ & $SU(3)^\prime$ & $SU(2)_L$ & $U(1)^\prime$ & $U(1)_{B^\prime-L^\prime}$ & $U(1)_\chi$ \\
                \hline                
                	$\chi$ & $\mathbf{15}$ & $\mathbf{1}$ & $\mathbf{1}$ & $0$ & $0$ & $-1$ \\	
                $F$  & $\mathbf{1}$ & $\mathbf{8}$ & $\mathbf{1}$ & $0$ & $-1$ & $0$ \\
		\hline
		\hline
		$\sigma_1$  & $\mathbf{1}$ & $\mathbf{1}$ & $\mathbf{1}$ & $0$ & $0$ & $2$ \\ 
                $\sigma_2$  & $\mathbf{1}$ & $\mathbf{1}$ & $\mathbf{1}$ & $0$ & $2$ & $0$ \\ 
		$\eta_{1,2}$ & $\mathbf{1}$ & $\mathbf{8}$ & $\mathbf{2}$ & $1/2$ & $0$ & $0$ \\	
		\hline
	\end{tabular}
	\caption{Scalars and fermions added to the model in~\cite{DiLuzio:2018zxy}. 
	\label{tab:bmlnur}}
\end{table}

\begin{figure}
\centering
\includegraphics[width=0.33\textwidth]{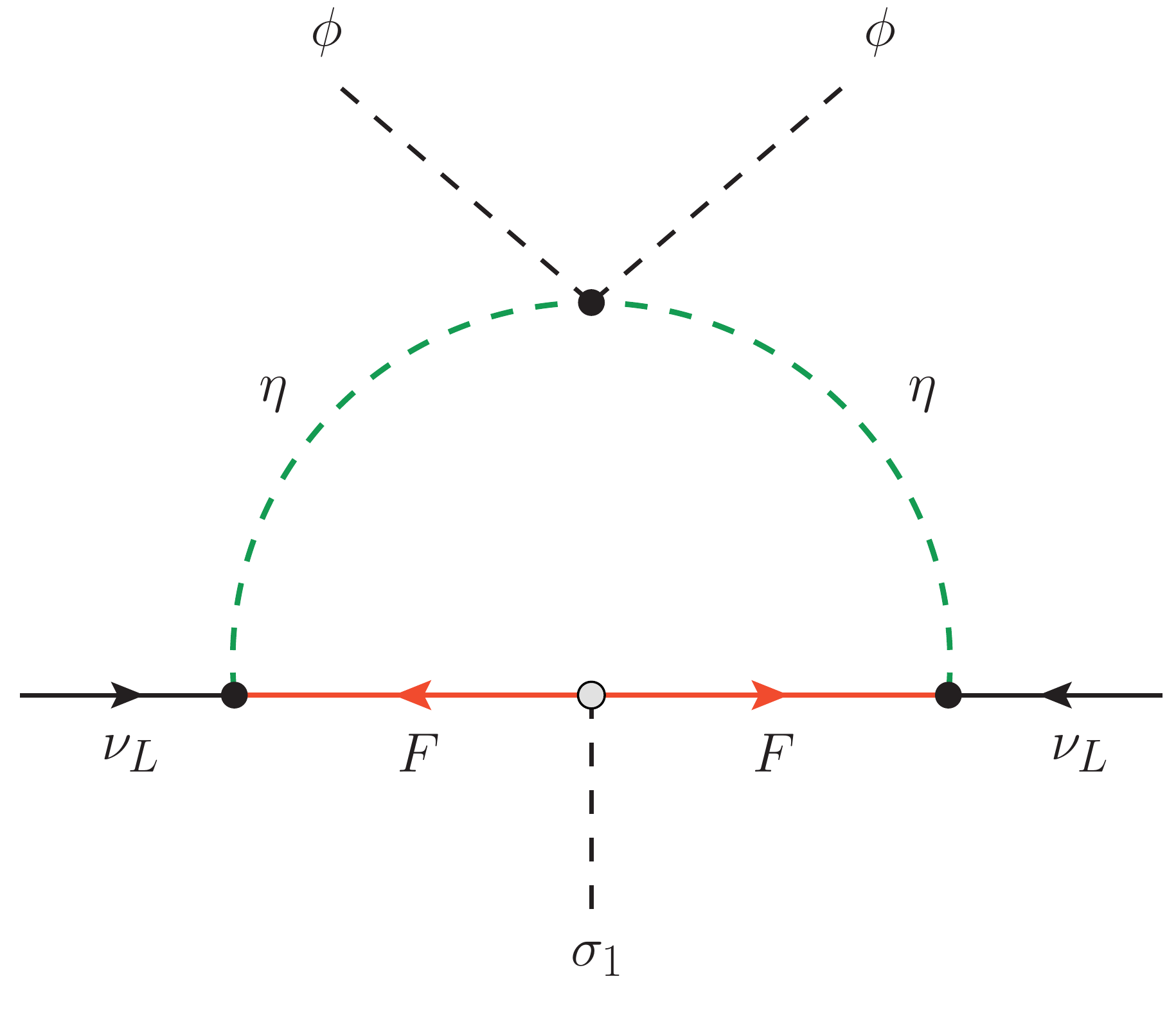}
\caption{\label{fig:scotogenicDiagram}
One-loop diagram for neutrino mass generation.}
\end{figure}

Thanks to the introduction of the fermion $F$ and the scalar fields $\eta_{1,2} \sim (\mathbf{1},\mathbf{8},\mathbf{2},1/2)$, the generation of neutrino masses takes place at the one-loop level \textit{\`a la scotogenic}~\cite{Ma:2006km}, in analogy to \cite{Ma:2017vdv,Reig:2018mdk}. The relevant Yukawa terms for the generation of neutrino masses are
\begin{equation} \label{eq:yukawa}
\mathcal{L}_Y \supset y_{i \alpha} \, \overline{F} \, \eta_i \, \ell_L^\alpha + \frac{y_F}{2} \, \sigma_2 \, \overline{F^c} \, F +\mathrm{h.c.}\, ,
\end{equation}
where $i=1,2$ and $\alpha=1,2,3$ are generation indices, while the scalar potential terms relevant for neutrino mass generation read
\begin{align}\label{eq:potential}
\begin{aligned}
V & \supset 
m_{\eta_i}^2 \, \eta_i^\dag \eta_i+\frac{1}{2} \, \lambda_\nu^{ij} \, \left[ \left( \phi^\dagger \, \eta_i \right)\left( \phi^\dagger \, \eta_j \right) + \, \hc \right] \,.
\end{aligned}
\end{align}
Here $\phi$ is the SM Higgs doublet with VEV $\langle \phi^0 \rangle = v / \sqrt{2}$. Without loss of generality, the mass matrix of the $\eta$ scalars has been chosen to be diagonal. In the presence of the Yukawa interactions in~\eqref{eq:yukawa} and the $\lambda_\nu$ term in~\eqref{eq:potential}, lepton number gets broken in two units. Assuming that the scalar potential is such that the $\eta_i$ scalars do not get VEVs, neutrino masses are forbidden at tree level and are generated via the loop diagram in Figure~\ref{fig:scotogenicDiagram}. Taking $F$ and $\eta$ masses close to the PQ breaking scale $v_{{\rm PQ}_2}$ and $\mathcal O(1)$ Yukawa couplings, neutrino masses are predicted to be $\sim \lambda_\nu v^2/(4 \pi^2 v_{{\rm PQ}_2})$, hence in the $\sim 0.1$ eV ballpark for $\lambda_\nu \in \left[10^{-6},10^{-1}\right]$ when $v_{{\rm PQ}_2} \in \left[10^8,10^{13}\right]$ GeV. In the limit $\lambda_\nu \ll 1$, the one-loop neutrino mass matrix reads
\begin{align}\label{eq:numass}
\left(m_{\nu}\right)_{\alpha\beta} =
\frac{v^2}{4 \, \pi^2 \, M_F} \, y_{i \alpha} \, y_{j \beta} \, \lambda_\nu^{kk} \, L_k \, X_{ijk}(\varphi) \, ,
\end{align}
where a sum over $i,j,k$ is implicit and where we have defined $M_F = y_F \, \langle \sigma_2 \rangle$, the loop function $L_i=\Sigma_i + \Sigma_i^2 \log [ M_F^2/m_{\eta_i}^2\, ]$, with $\Sigma_i=M_F^2/(m_{\eta_i}^2 - M_F^2)$, and
\begin{align}
\begin{aligned}
X_{ijk}(\varphi) &= \delta_{ij} \left[ \sin^2 \varphi + \left(\cos^2\varphi - \sin^2\varphi\right) \, \delta_{jk} \right]\\
&\quad + \left(1-\delta_{ij}\right) \left(\delta_{k2} - \delta_{k1}\right) \, \sin\varphi \cos\varphi \,,
\end{aligned}
\end{align}
with $\varphi$ being the $\eta_1 - \eta_2$ mixing angle. Since we included only two $\eta$ scalars (and a unique $F$), our framework predicts one massless neutrino. Such a minimal setup was also considered in \cite{Hehn:2012kz}.

Finally, one might be concerned about the stability of the neutral component of the $\chi$ multiplet. In the pre-inflationary case, however, the cosmic abundance of this specie (with mass typically of the order of the PQ scale) will be diluted due to the effect of cosmic inflation.

\section{Discussion}
Among the different options proposed to explain the hints of LFUV observed in $B$-meson decays, the hypothesis of $U_1$ vector leptoquark stands for its simplicity and effectiveness.  The requirement of a phenomenologically consistent TeV scale $U_1$ arising from an extended gauge group implies the emergence of QCD interactions from a nontrivial high-energy group structure. The solution of the strong CP problem in this class of models {\`a} la Peccei-Quinn predicts the existence of two axions. Remarkably, many of the properties of this extended axion sector are specified in terms of parameters that can be deduced from the explanation of the LFUV hints. The exploration of axion signatures thus offers a complementary probe of the possible underlying NP structure behind by these deviations. As an example of this complementarity, we have shown that post-inflationary axions are strongly disfavored in these setups. A better measurement of the isocurvature fluctuations in the cosmic-microwave background could therefore shed some light on the viability of these models. 

Many of the findings of this article go beyond the hints of LFUV in $B$-meson decays and can be generally applied to a large class of models where QCD arises as a low-energy interaction of an extended gauge sector. As we have shown, constraints that are typically less relevant for the QCD axion, such as those from X-rays or EBL, can provide crucial information in this class of models. Moreover, the extended axion sector can account for the observed DM relic abundance outside the QCD band. 

These predictions strengthen the motivation for the search of axion-like particles solving the strong CP problem in parameter space regions that extend beyond that of the QCD axion.



\acknowledgements
We thank Gino Isidori for carefully reading the manuscript and for useful comments. We are also grateful to Prateek Agrawal and Alejandro Ibarra for fruitful discussions. We thank the CERN Theory Department for hospitality while the original idea for this manuscript was conceived. M.R. thanks Pablo Quilez and Rachel Houtz for useful discussions. The work of M.R. and A.V. was supported by the Spanish grants SEV-2014-0398 and FPA2017-85216-P (AEI/FEDER, UE) and SEJI/2018/033 (Generalitat Valenciana) and the Spanish Red Consolider MultiDark FPA2017-90566-REDC. The work of J.F. was supported in part by the Swiss National Science Foundation (SNF) under contract 200021-159720 and by the Generalitat Valenciana under contract SEJI/2018/033.

\bibliography{references}

\begin{thebibliography}{64}%
\makeatletter
\providecommand \@ifxundefined [1]{%
 \@ifx{#1\undefined}
}%
\providecommand \@ifnum [1]{%
 \ifnum #1\expandafter \@firstoftwo
 \else \expandafter \@secondoftwo
 \fi
}%
\providecommand \@ifx [1]{%
 \ifx #1\expandafter \@firstoftwo
 \else \expandafter \@secondoftwo
 \fi
}%
\providecommand \natexlab [1]{#1}%
\providecommand \enquote  [1]{``#1''}%
\providecommand \bibnamefont  [1]{#1}%
\providecommand \bibfnamefont [1]{#1}%
\providecommand \citenamefont [1]{#1}%
\providecommand \href@noop [0]{\@secondoftwo}%
\providecommand \href [0]{\begingroup \@sanitize@url \@href}%
\providecommand \@href[1]{\@@startlink{#1}\@@href}%
\providecommand \@@href[1]{\endgroup#1\@@endlink}%
\providecommand \@sanitize@url [0]{\catcode `\\12\catcode `\$12\catcode
  `\&12\catcode `\#12\catcode `\^12\catcode `\_12\catcode `\%12\relax}%
\providecommand \@@startlink[1]{}%
\providecommand \@@endlink[0]{}%
\providecommand \url  [0]{\begingroup\@sanitize@url \@url }%
\providecommand \@url [1]{\endgroup\@href {#1}{\urlprefix }}%
\providecommand \urlprefix  [0]{URL }%
\providecommand \Eprint [0]{\href }%
\providecommand \doibase [0]{http://dx.doi.org/}%
\providecommand \selectlanguage [0]{\@gobble}%
\providecommand \bibinfo  [0]{\@secondoftwo}%
\providecommand \bibfield  [0]{\@secondoftwo}%
\providecommand \translation [1]{[#1]}%
\providecommand \BibitemOpen [0]{}%
\providecommand \bibitemStop [0]{}%
\providecommand \bibitemNoStop [0]{.\EOS\space}%
\providecommand \EOS [0]{\spacefactor3000\relax}%
\providecommand \BibitemShut  [1]{\csname bibitem#1\endcsname}%
\let\auto@bib@innerbib\@empty
\bibitem [{\citenamefont {Lees}\ \emph {et~al.}(2013)\citenamefont {Lees} \emph
  {et~al.}}]{Lees:2013uzd}%
  \BibitemOpen
  \bibfield  {author} {\bibinfo {author} {\bibfnamefont {J.~P.}\ \bibnamefont
  {Lees}} \emph {et~al.} (\bibinfo {collaboration} {BaBar}),\ }\href {\doibase
  10.1103/PhysRevD.88.072012} {\bibfield  {journal} {\bibinfo  {journal} {Phys.
  Rev.}\ }\textbf {\bibinfo {volume} {D88}},\ \bibinfo {pages} {072012}
  (\bibinfo {year} {2013})},\ \Eprint {http://arxiv.org/abs/1303.0571}
  {arXiv:1303.0571 [hep-ex]} \BibitemShut {NoStop}%
\bibitem [{\citenamefont {Aaij}\ \emph {et~al.}(2015)\citenamefont {Aaij} \emph
  {et~al.}}]{Aaij:2015yra}%
  \BibitemOpen
  \bibfield  {author} {\bibinfo {author} {\bibfnamefont {R.}~\bibnamefont
  {Aaij}} \emph {et~al.} (\bibinfo {collaboration} {LHCb}),\ }\href {\doibase
  10.1103/PhysRevLett.115.159901, 10.1103/PhysRevLett.115.111803} {\bibfield
  {journal} {\bibinfo  {journal} {Phys. Rev. Lett.}\ }\textbf {\bibinfo
  {volume} {115}},\ \bibinfo {pages} {111803} (\bibinfo {year} {2015})},\
  \bibinfo {note} {[Erratum: Phys. Rev. Lett.115,no.15,159901(2015)]},\ \Eprint
  {http://arxiv.org/abs/1506.08614} {arXiv:1506.08614 [hep-ex]} \BibitemShut
  {NoStop}%
\bibitem [{\citenamefont {Hirose}\ \emph {et~al.}(2017)\citenamefont {Hirose}
  \emph {et~al.}}]{Hirose:2016wfn}%
  \BibitemOpen
  \bibfield  {author} {\bibinfo {author} {\bibfnamefont {S.}~\bibnamefont
  {Hirose}} \emph {et~al.} (\bibinfo {collaboration} {Belle}),\ }\href
  {\doibase 10.1103/PhysRevLett.118.211801} {\bibfield  {journal} {\bibinfo
  {journal} {Phys. Rev. Lett.}\ }\textbf {\bibinfo {volume} {118}},\ \bibinfo
  {pages} {211801} (\bibinfo {year} {2017})},\ \Eprint
  {http://arxiv.org/abs/1612.00529} {arXiv:1612.00529 [hep-ex]} \BibitemShut
  {NoStop}%
\bibitem [{\citenamefont {Aaij}\ \emph {et~al.}(2018)\citenamefont {Aaij} \emph
  {et~al.}}]{Aaij:2017deq}%
  \BibitemOpen
  \bibfield  {author} {\bibinfo {author} {\bibfnamefont {R.}~\bibnamefont
  {Aaij}} \emph {et~al.} (\bibinfo {collaboration} {LHCb}),\ }\href {\doibase
  10.1103/PhysRevD.97.072013} {\bibfield  {journal} {\bibinfo  {journal} {Phys.
  Rev.}\ }\textbf {\bibinfo {volume} {D97}},\ \bibinfo {pages} {072013}
  (\bibinfo {year} {2018})},\ \Eprint {http://arxiv.org/abs/1711.02505}
  {arXiv:1711.02505 [hep-ex]} \BibitemShut {NoStop}%
\bibitem [{\citenamefont {Abdesselam}\ \emph
  {et~al.}(2019{\natexlab{a}})\citenamefont {Abdesselam} \emph
  {et~al.}}]{Abdesselam:2019dgh}%
  \BibitemOpen
  \bibfield  {author} {\bibinfo {author} {\bibfnamefont {A.}~\bibnamefont
  {Abdesselam}} \emph {et~al.} (\bibinfo {collaboration} {Belle}),\ }\href@noop
  {} {\  (\bibinfo {year} {2019}{\natexlab{a}})},\ \Eprint
  {http://arxiv.org/abs/1904.08794} {arXiv:1904.08794 [hep-ex]} \BibitemShut
  {NoStop}%
\bibitem [{\citenamefont {Aaij}\ \emph {et~al.}(2014)\citenamefont {Aaij} \emph
  {et~al.}}]{Aaij:2014ora}%
  \BibitemOpen
  \bibfield  {author} {\bibinfo {author} {\bibfnamefont {R.}~\bibnamefont
  {Aaij}} \emph {et~al.} (\bibinfo {collaboration} {LHCb}),\ }\href {\doibase
  10.1103/PhysRevLett.113.151601} {\bibfield  {journal} {\bibinfo  {journal}
  {Phys. Rev. Lett.}\ }\textbf {\bibinfo {volume} {113}},\ \bibinfo {pages}
  {151601} (\bibinfo {year} {2014})},\ \Eprint {http://arxiv.org/abs/1406.6482}
  {arXiv:1406.6482 [hep-ex]} \BibitemShut {NoStop}%
\bibitem [{\citenamefont {Aaij}\ \emph {et~al.}(2017)\citenamefont {Aaij} \emph
  {et~al.}}]{Aaij:2017vbb}%
  \BibitemOpen
  \bibfield  {author} {\bibinfo {author} {\bibfnamefont {R.}~\bibnamefont
  {Aaij}} \emph {et~al.} (\bibinfo {collaboration} {LHCb}),\ }\href {\doibase
  10.1007/JHEP08(2017)055} {\bibfield  {journal} {\bibinfo  {journal} {JHEP}\
  }\textbf {\bibinfo {volume} {08}},\ \bibinfo {pages} {055} (\bibinfo {year}
  {2017})},\ \Eprint {http://arxiv.org/abs/1705.05802} {arXiv:1705.05802
  [hep-ex]} \BibitemShut {NoStop}%
\bibitem [{\citenamefont {Aaij}\ \emph {et~al.}(2019)\citenamefont {Aaij} \emph
  {et~al.}}]{Aaij:2019wad}%
  \BibitemOpen
  \bibfield  {author} {\bibinfo {author} {\bibfnamefont {R.}~\bibnamefont
  {Aaij}} \emph {et~al.} (\bibinfo {collaboration} {LHCb}),\ }\href {\doibase
  10.1103/PhysRevLett.122.191801} {\bibfield  {journal} {\bibinfo  {journal}
  {Phys. Rev. Lett.}\ }\textbf {\bibinfo {volume} {122}},\ \bibinfo {pages}
  {191801} (\bibinfo {year} {2019})},\ \Eprint
  {http://arxiv.org/abs/1903.09252} {arXiv:1903.09252 [hep-ex]} \BibitemShut
  {NoStop}%
\bibitem [{\citenamefont {Abdesselam}\ \emph
  {et~al.}(2019{\natexlab{b}})\citenamefont {Abdesselam} \emph
  {et~al.}}]{Abdesselam:2019wac}%
  \BibitemOpen
  \bibfield  {author} {\bibinfo {author} {\bibfnamefont {A.}~\bibnamefont
  {Abdesselam}} \emph {et~al.} (\bibinfo {collaboration} {Belle}),\ }\href@noop
  {} {\  (\bibinfo {year} {2019}{\natexlab{b}})},\ \Eprint
  {http://arxiv.org/abs/1904.02440} {arXiv:1904.02440 [hep-ex]} \BibitemShut
  {NoStop}%
\bibitem [{\citenamefont {Algueró}\ \emph {et~al.}(2019)\citenamefont
  {Algueró}, \citenamefont {Capdevila}, \citenamefont {Crivellin},
  \citenamefont {Descotes-Genon}, \citenamefont {Masjuan}, \citenamefont
  {Matias},\ and\ \citenamefont {Virto}}]{Alguero:2019ptt}%
  \BibitemOpen
  \bibfield  {author} {\bibinfo {author} {\bibfnamefont {M.}~\bibnamefont
  {Algueró}}, \bibinfo {author} {\bibfnamefont {B.}~\bibnamefont {Capdevila}},
  \bibinfo {author} {\bibfnamefont {A.}~\bibnamefont {Crivellin}}, \bibinfo
  {author} {\bibfnamefont {S.}~\bibnamefont {Descotes-Genon}}, \bibinfo
  {author} {\bibfnamefont {P.}~\bibnamefont {Masjuan}}, \bibinfo {author}
  {\bibfnamefont {J.}~\bibnamefont {Matias}}, \ and\ \bibinfo {author}
  {\bibfnamefont {J.}~\bibnamefont {Virto}},\ }\href {\doibase
  10.1140/epjc/s10052-019-7216-3} {\bibfield  {journal} {\bibinfo  {journal}
  {Eur. Phys. J.}\ }\textbf {\bibinfo {volume} {C79}},\ \bibinfo {pages} {714}
  (\bibinfo {year} {2019})},\ \Eprint {http://arxiv.org/abs/1903.09578}
  {arXiv:1903.09578 [hep-ph]} \BibitemShut {NoStop}%
\bibitem [{\citenamefont {Aebischer}\ \emph {et~al.}(2019)\citenamefont
  {Aebischer}, \citenamefont {Altmannshofer}, \citenamefont {Guadagnoli},
  \citenamefont {Reboud}, \citenamefont {Stangl},\ and\ \citenamefont
  {Straub}}]{Aebischer:2019mlg}%
  \BibitemOpen
  \bibfield  {author} {\bibinfo {author} {\bibfnamefont {J.}~\bibnamefont
  {Aebischer}}, \bibinfo {author} {\bibfnamefont {W.}~\bibnamefont
  {Altmannshofer}}, \bibinfo {author} {\bibfnamefont {D.}~\bibnamefont
  {Guadagnoli}}, \bibinfo {author} {\bibfnamefont {M.}~\bibnamefont {Reboud}},
  \bibinfo {author} {\bibfnamefont {P.}~\bibnamefont {Stangl}}, \ and\ \bibinfo
  {author} {\bibfnamefont {D.~M.}\ \bibnamefont {Straub}},\ }\href@noop {} {\
  (\bibinfo {year} {2019})},\ \Eprint {http://arxiv.org/abs/1903.10434}
  {arXiv:1903.10434 [hep-ph]} \BibitemShut {NoStop}%
\bibitem [{\citenamefont {Ciuchini}\ \emph {et~al.}(2019)\citenamefont
  {Ciuchini}, \citenamefont {Coutinho}, \citenamefont {Fedele}, \citenamefont
  {Franco}, \citenamefont {Paul}, \citenamefont {Silvestrini},\ and\
  \citenamefont {Valli}}]{Ciuchini:2019usw}%
  \BibitemOpen
  \bibfield  {author} {\bibinfo {author} {\bibfnamefont {M.}~\bibnamefont
  {Ciuchini}}, \bibinfo {author} {\bibfnamefont {A.~M.}\ \bibnamefont
  {Coutinho}}, \bibinfo {author} {\bibfnamefont {M.}~\bibnamefont {Fedele}},
  \bibinfo {author} {\bibfnamefont {E.}~\bibnamefont {Franco}}, \bibinfo
  {author} {\bibfnamefont {A.}~\bibnamefont {Paul}}, \bibinfo {author}
  {\bibfnamefont {L.}~\bibnamefont {Silvestrini}}, \ and\ \bibinfo {author}
  {\bibfnamefont {M.}~\bibnamefont {Valli}},\ }\href {\doibase
  10.1140/epjc/s10052-019-7210-9} {\bibfield  {journal} {\bibinfo  {journal}
  {Eur. Phys. J.}\ }\textbf {\bibinfo {volume} {C79}},\ \bibinfo {pages} {719}
  (\bibinfo {year} {2019})},\ \Eprint {http://arxiv.org/abs/1903.09632}
  {arXiv:1903.09632 [hep-ph]} \BibitemShut {NoStop}%
\bibitem [{\citenamefont {Datta}\ \emph {et~al.}(2019)\citenamefont {Datta},
  \citenamefont {Kumar},\ and\ \citenamefont {London}}]{Datta:2019zca}%
  \BibitemOpen
  \bibfield  {author} {\bibinfo {author} {\bibfnamefont {A.}~\bibnamefont
  {Datta}}, \bibinfo {author} {\bibfnamefont {J.}~\bibnamefont {Kumar}}, \ and\
  \bibinfo {author} {\bibfnamefont {D.}~\bibnamefont {London}},\ }\href
  {\doibase 10.1016/j.physletb.2019.134858} {\bibfield  {journal} {\bibinfo
  {journal} {Phys. Lett.}\ }\textbf {\bibinfo {volume} {B797}},\ \bibinfo
  {pages} {134858} (\bibinfo {year} {2019})},\ \Eprint
  {http://arxiv.org/abs/1903.10086} {arXiv:1903.10086 [hep-ph]} \BibitemShut
  {NoStop}%
\bibitem [{\citenamefont {Shi}\ \emph {et~al.}(2019)\citenamefont {Shi},
  \citenamefont {Geng}, \citenamefont {Grinstein}, \citenamefont {Jäger},\
  and\ \citenamefont {Martin~Camalich}}]{Shi:2019gxi}%
  \BibitemOpen
  \bibfield  {author} {\bibinfo {author} {\bibfnamefont {R.-X.}\ \bibnamefont
  {Shi}}, \bibinfo {author} {\bibfnamefont {L.-S.}\ \bibnamefont {Geng}},
  \bibinfo {author} {\bibfnamefont {B.}~\bibnamefont {Grinstein}}, \bibinfo
  {author} {\bibfnamefont {S.}~\bibnamefont {Jäger}}, \ and\ \bibinfo {author}
  {\bibfnamefont {J.}~\bibnamefont {Martin~Camalich}},\ }\href {\doibase
  10.1007/JHEP12(2019)065} {\bibfield  {journal} {\bibinfo  {journal} {JHEP}\
  }\textbf {\bibinfo {volume} {12}},\ \bibinfo {pages} {065} (\bibinfo {year}
  {2019})},\ \Eprint {http://arxiv.org/abs/1905.08498} {arXiv:1905.08498
  [hep-ph]} \BibitemShut {NoStop}%
\bibitem [{\citenamefont {Murgui}\ \emph {et~al.}(2019)\citenamefont {Murgui},
  \citenamefont {Peñuelas}, \citenamefont {Jung},\ and\ \citenamefont
  {Pich}}]{Murgui:2019czp}%
  \BibitemOpen
  \bibfield  {author} {\bibinfo {author} {\bibfnamefont {C.}~\bibnamefont
  {Murgui}}, \bibinfo {author} {\bibfnamefont {A.}~\bibnamefont {Peñuelas}},
  \bibinfo {author} {\bibfnamefont {M.}~\bibnamefont {Jung}}, \ and\ \bibinfo
  {author} {\bibfnamefont {A.}~\bibnamefont {Pich}},\ }\href {\doibase
  10.1007/JHEP09(2019)103} {\bibfield  {journal} {\bibinfo  {journal} {JHEP}\
  }\textbf {\bibinfo {volume} {09}},\ \bibinfo {pages} {103} (\bibinfo {year}
  {2019})},\ \Eprint {http://arxiv.org/abs/1904.09311} {arXiv:1904.09311
  [hep-ph]} \BibitemShut {NoStop}%
\bibitem [{\citenamefont {Barbieri}\ \emph {et~al.}(2016)\citenamefont
  {Barbieri}, \citenamefont {Isidori}, \citenamefont {Pattori},\ and\
  \citenamefont {Senia}}]{Barbieri:2015yvd}%
  \BibitemOpen
  \bibfield  {author} {\bibinfo {author} {\bibfnamefont {R.}~\bibnamefont
  {Barbieri}}, \bibinfo {author} {\bibfnamefont {G.}~\bibnamefont {Isidori}},
  \bibinfo {author} {\bibfnamefont {A.}~\bibnamefont {Pattori}}, \ and\
  \bibinfo {author} {\bibfnamefont {F.}~\bibnamefont {Senia}},\ }\href
  {\doibase 10.1140/epjc/s10052-016-3905-3} {\bibfield  {journal} {\bibinfo
  {journal} {Eur. Phys. J.}\ }\textbf {\bibinfo {volume} {C76}},\ \bibinfo
  {pages} {67} (\bibinfo {year} {2016})},\ \Eprint
  {http://arxiv.org/abs/1512.01560} {arXiv:1512.01560 [hep-ph]} \BibitemShut
  {NoStop}%
\bibitem [{\citenamefont {Buttazzo}\ \emph {et~al.}(2017)\citenamefont
  {Buttazzo}, \citenamefont {Greljo}, \citenamefont {Isidori},\ and\
  \citenamefont {Marzocca}}]{Buttazzo:2017ixm}%
  \BibitemOpen
  \bibfield  {author} {\bibinfo {author} {\bibfnamefont {D.}~\bibnamefont
  {Buttazzo}}, \bibinfo {author} {\bibfnamefont {A.}~\bibnamefont {Greljo}},
  \bibinfo {author} {\bibfnamefont {G.}~\bibnamefont {Isidori}}, \ and\
  \bibinfo {author} {\bibfnamefont {D.}~\bibnamefont {Marzocca}},\ }\href
  {\doibase 10.1007/JHEP11(2017)044} {\bibfield  {journal} {\bibinfo  {journal}
  {JHEP}\ }\textbf {\bibinfo {volume} {11}},\ \bibinfo {pages} {044} (\bibinfo
  {year} {2017})},\ \Eprint {http://arxiv.org/abs/1706.07808} {arXiv:1706.07808
  [hep-ph]} \BibitemShut {NoStop}%
\bibitem [{\citenamefont {Barbieri}\ and\ \citenamefont
  {Ziegler}(2019)}]{Barbieri:2019zdz}%
  \BibitemOpen
  \bibfield  {author} {\bibinfo {author} {\bibfnamefont {R.}~\bibnamefont
  {Barbieri}}\ and\ \bibinfo {author} {\bibfnamefont {R.}~\bibnamefont
  {Ziegler}},\ }\href {\doibase 10.1007/JHEP07(2019)023} {\bibfield  {journal}
  {\bibinfo  {journal} {JHEP}\ }\textbf {\bibinfo {volume} {07}},\ \bibinfo
  {pages} {023} (\bibinfo {year} {2019})},\ \Eprint
  {http://arxiv.org/abs/1904.04121} {arXiv:1904.04121 [hep-ph]} \BibitemShut
  {NoStop}%
\bibitem [{\citenamefont {Bordone}\ \emph
  {et~al.}(2018{\natexlab{a}})\citenamefont {Bordone}, \citenamefont
  {Cornella}, \citenamefont {Fuentes-Martin},\ and\ \citenamefont
  {Isidori}}]{Bordone:2017bld}%
  \BibitemOpen
  \bibfield  {author} {\bibinfo {author} {\bibfnamefont {M.}~\bibnamefont
  {Bordone}}, \bibinfo {author} {\bibfnamefont {C.}~\bibnamefont {Cornella}},
  \bibinfo {author} {\bibfnamefont {J.}~\bibnamefont {Fuentes-Martin}}, \ and\
  \bibinfo {author} {\bibfnamefont {G.}~\bibnamefont {Isidori}},\ }\href
  {\doibase 10.1016/j.physletb.2018.02.011} {\bibfield  {journal} {\bibinfo
  {journal} {Phys. Lett.}\ }\textbf {\bibinfo {volume} {B779}},\ \bibinfo
  {pages} {317} (\bibinfo {year} {2018}{\natexlab{a}})},\ \Eprint
  {http://arxiv.org/abs/1712.01368} {arXiv:1712.01368 [hep-ph]} \BibitemShut
  {NoStop}%
\bibitem [{\citenamefont {Greljo}\ and\ \citenamefont
  {Stefanek}(2018)}]{Greljo:2018tuh}%
  \BibitemOpen
  \bibfield  {author} {\bibinfo {author} {\bibfnamefont {A.}~\bibnamefont
  {Greljo}}\ and\ \bibinfo {author} {\bibfnamefont {B.~A.}\ \bibnamefont
  {Stefanek}},\ }\href {\doibase 10.1016/j.physletb.2018.05.033} {\bibfield
  {journal} {\bibinfo  {journal} {Phys. Lett.}\ }\textbf {\bibinfo {volume}
  {B782}},\ \bibinfo {pages} {131} (\bibinfo {year} {2018})},\ \Eprint
  {http://arxiv.org/abs/1802.04274} {arXiv:1802.04274 [hep-ph]} \BibitemShut
  {NoStop}%
\bibitem [{\citenamefont {Bordone}\ \emph
  {et~al.}(2018{\natexlab{b}})\citenamefont {Bordone}, \citenamefont
  {Cornella}, \citenamefont {Fuentes-Martín},\ and\ \citenamefont
  {Isidori}}]{Bordone:2018nbg}%
  \BibitemOpen
  \bibfield  {author} {\bibinfo {author} {\bibfnamefont {M.}~\bibnamefont
  {Bordone}}, \bibinfo {author} {\bibfnamefont {C.}~\bibnamefont {Cornella}},
  \bibinfo {author} {\bibfnamefont {J.}~\bibnamefont {Fuentes-Martín}}, \ and\
  \bibinfo {author} {\bibfnamefont {G.}~\bibnamefont {Isidori}},\ }\href
  {\doibase 10.1007/JHEP10(2018)148} {\bibfield  {journal} {\bibinfo  {journal}
  {JHEP}\ }\textbf {\bibinfo {volume} {10}},\ \bibinfo {pages} {148} (\bibinfo
  {year} {2018}{\natexlab{b}})},\ \Eprint {http://arxiv.org/abs/1805.09328}
  {arXiv:1805.09328 [hep-ph]} \BibitemShut {NoStop}%
\bibitem [{\citenamefont {Cornella}\ \emph {et~al.}(2019)\citenamefont
  {Cornella}, \citenamefont {Fuentes-Martin},\ and\ \citenamefont
  {Isidori}}]{Cornella:2019hct}%
  \BibitemOpen
  \bibfield  {author} {\bibinfo {author} {\bibfnamefont {C.}~\bibnamefont
  {Cornella}}, \bibinfo {author} {\bibfnamefont {J.}~\bibnamefont
  {Fuentes-Martin}}, \ and\ \bibinfo {author} {\bibfnamefont {G.}~\bibnamefont
  {Isidori}},\ }\href {\doibase 10.1007/JHEP07(2019)168} {\bibfield  {journal}
  {\bibinfo  {journal} {JHEP}\ }\textbf {\bibinfo {volume} {07}},\ \bibinfo
  {pages} {168} (\bibinfo {year} {2019})},\ \Eprint
  {http://arxiv.org/abs/1903.11517} {arXiv:1903.11517 [hep-ph]} \BibitemShut
  {NoStop}%
\bibitem [{\citenamefont {Georgi}\ and\ \citenamefont
  {Nakai}(2016)}]{Georgi:2016xhm}%
  \BibitemOpen
  \bibfield  {author} {\bibinfo {author} {\bibfnamefont {H.}~\bibnamefont
  {Georgi}}\ and\ \bibinfo {author} {\bibfnamefont {Y.}~\bibnamefont {Nakai}},\
  }\href {\doibase 10.1103/PhysRevD.94.075005} {\bibfield  {journal} {\bibinfo
  {journal} {Phys. Rev.}\ }\textbf {\bibinfo {volume} {D94}},\ \bibinfo {pages}
  {075005} (\bibinfo {year} {2016})},\ \Eprint
  {http://arxiv.org/abs/1606.05865} {arXiv:1606.05865 [hep-ph]} \BibitemShut
  {NoStop}%
\bibitem [{\citenamefont {Diaz}\ \emph {et~al.}(2017)\citenamefont {Diaz},
  \citenamefont {Schmaltz},\ and\ \citenamefont {Zhong}}]{Diaz:2017lit}%
  \BibitemOpen
  \bibfield  {author} {\bibinfo {author} {\bibfnamefont {B.}~\bibnamefont
  {Diaz}}, \bibinfo {author} {\bibfnamefont {M.}~\bibnamefont {Schmaltz}}, \
  and\ \bibinfo {author} {\bibfnamefont {Y.-M.}\ \bibnamefont {Zhong}},\ }\href
  {\doibase 10.1007/JHEP10(2017)097} {\bibfield  {journal} {\bibinfo  {journal}
  {JHEP}\ }\textbf {\bibinfo {volume} {10}},\ \bibinfo {pages} {097} (\bibinfo
  {year} {2017})},\ \Eprint {http://arxiv.org/abs/1706.05033} {arXiv:1706.05033
  [hep-ph]} \BibitemShut {NoStop}%
\bibitem [{\citenamefont {Di~Luzio}\ \emph {et~al.}(2017)\citenamefont
  {Di~Luzio}, \citenamefont {Greljo},\ and\ \citenamefont
  {Nardecchia}}]{DiLuzio:2017vat}%
  \BibitemOpen
  \bibfield  {author} {\bibinfo {author} {\bibfnamefont {L.}~\bibnamefont
  {Di~Luzio}}, \bibinfo {author} {\bibfnamefont {A.}~\bibnamefont {Greljo}}, \
  and\ \bibinfo {author} {\bibfnamefont {M.}~\bibnamefont {Nardecchia}},\
  }\href {\doibase 10.1103/PhysRevD.96.115011} {\bibfield  {journal} {\bibinfo
  {journal} {Phys. Rev.}\ }\textbf {\bibinfo {volume} {D96}},\ \bibinfo {pages}
  {115011} (\bibinfo {year} {2017})},\ \Eprint
  {http://arxiv.org/abs/1708.08450} {arXiv:1708.08450 [hep-ph]} \BibitemShut
  {NoStop}%
\bibitem [{\citenamefont {Blanke}\ and\ \citenamefont
  {Crivellin}(2018)}]{Blanke:2018sro}%
  \BibitemOpen
  \bibfield  {author} {\bibinfo {author} {\bibfnamefont {M.}~\bibnamefont
  {Blanke}}\ and\ \bibinfo {author} {\bibfnamefont {A.}~\bibnamefont
  {Crivellin}},\ }\href {\doibase 10.1103/PhysRevLett.121.011801} {\bibfield
  {journal} {\bibinfo  {journal} {Phys. Rev. Lett.}\ }\textbf {\bibinfo
  {volume} {121}},\ \bibinfo {pages} {011801} (\bibinfo {year} {2018})},\
  \Eprint {http://arxiv.org/abs/1801.07256} {arXiv:1801.07256 [hep-ph]}
  \BibitemShut {NoStop}%
\bibitem [{\citenamefont {Di~Luzio}\ \emph {et~al.}(2018)\citenamefont
  {Di~Luzio}, \citenamefont {Fuentes-Martin}, \citenamefont {Greljo},
  \citenamefont {Nardecchia},\ and\ \citenamefont {Renner}}]{DiLuzio:2018zxy}%
  \BibitemOpen
  \bibfield  {author} {\bibinfo {author} {\bibfnamefont {L.}~\bibnamefont
  {Di~Luzio}}, \bibinfo {author} {\bibfnamefont {J.}~\bibnamefont
  {Fuentes-Martin}}, \bibinfo {author} {\bibfnamefont {A.}~\bibnamefont
  {Greljo}}, \bibinfo {author} {\bibfnamefont {M.}~\bibnamefont {Nardecchia}},
  \ and\ \bibinfo {author} {\bibfnamefont {S.}~\bibnamefont {Renner}},\ }\href
  {\doibase 10.1007/JHEP11(2018)081} {\bibfield  {journal} {\bibinfo  {journal}
  {JHEP}\ }\textbf {\bibinfo {volume} {11}},\ \bibinfo {pages} {081} (\bibinfo
  {year} {2018})},\ \Eprint {http://arxiv.org/abs/1808.00942} {arXiv:1808.00942
  [hep-ph]} \BibitemShut {NoStop}%
\bibitem [{\citenamefont {Baker}\ \emph {et~al.}(2019)\citenamefont {Baker},
  \citenamefont {Fuentes-Martín}, \citenamefont {Isidori},\ and\ \citenamefont
  {König}}]{Baker:2019sli}%
  \BibitemOpen
  \bibfield  {author} {\bibinfo {author} {\bibfnamefont {M.~J.}\ \bibnamefont
  {Baker}}, \bibinfo {author} {\bibfnamefont {J.}~\bibnamefont
  {Fuentes-Martín}}, \bibinfo {author} {\bibfnamefont {G.}~\bibnamefont
  {Isidori}}, \ and\ \bibinfo {author} {\bibfnamefont {M.}~\bibnamefont
  {König}},\ }\href {\doibase 10.1140/epjc/s10052-019-6853-x} {\bibfield
  {journal} {\bibinfo  {journal} {Eur. Phys. J.}\ }\textbf {\bibinfo {volume}
  {C79}},\ \bibinfo {pages} {334} (\bibinfo {year} {2019})},\ \Eprint
  {http://arxiv.org/abs/1901.10480} {arXiv:1901.10480 [hep-ph]} \BibitemShut
  {NoStop}%
\bibitem [{\citenamefont {Ellis}\ and\ \citenamefont
  {Gaillard}(1979)}]{Ellis:1978hq}%
  \BibitemOpen
  \bibfield  {author} {\bibinfo {author} {\bibfnamefont {J.~R.}\ \bibnamefont
  {Ellis}}\ and\ \bibinfo {author} {\bibfnamefont {M.~K.}\ \bibnamefont
  {Gaillard}},\ }\href {\doibase 10.1016/0550-3213(79)90297-9} {\bibfield
  {journal} {\bibinfo  {journal} {Nucl. Phys.}\ }\textbf {\bibinfo {volume}
  {B150}},\ \bibinfo {pages} {141} (\bibinfo {year} {1979})}\BibitemShut
  {NoStop}%
\bibitem [{\citenamefont {Dugan}\ \emph {et~al.}(1985)\citenamefont {Dugan},
  \citenamefont {Grinstein},\ and\ \citenamefont {Hall}}]{Dugan:1984qf}%
  \BibitemOpen
  \bibfield  {author} {\bibinfo {author} {\bibfnamefont {M.}~\bibnamefont
  {Dugan}}, \bibinfo {author} {\bibfnamefont {B.}~\bibnamefont {Grinstein}}, \
  and\ \bibinfo {author} {\bibfnamefont {L.~J.}\ \bibnamefont {Hall}},\ }\href
  {\doibase 10.1016/0550-3213(85)90145-2} {\bibfield  {journal} {\bibinfo
  {journal} {Nucl. Phys.}\ }\textbf {\bibinfo {volume} {B255}},\ \bibinfo
  {pages} {413} (\bibinfo {year} {1985})}\BibitemShut {NoStop}%
\bibitem [{\citenamefont {Agrawal}\ and\ \citenamefont
  {Howe}(2018{\natexlab{a}})}]{Agrawal:2017ksf}%
  \BibitemOpen
  \bibfield  {author} {\bibinfo {author} {\bibfnamefont {P.}~\bibnamefont
  {Agrawal}}\ and\ \bibinfo {author} {\bibfnamefont {K.}~\bibnamefont {Howe}},\
  }\href {\doibase 10.1007/JHEP12(2018)029} {\bibfield  {journal} {\bibinfo
  {journal} {JHEP}\ }\textbf {\bibinfo {volume} {12}},\ \bibinfo {pages} {029}
  (\bibinfo {year} {2018}{\natexlab{a}})},\ \Eprint
  {http://arxiv.org/abs/1710.04213} {arXiv:1710.04213 [hep-ph]} \BibitemShut
  {NoStop}%
\bibitem [{\citenamefont {Agrawal}\ and\ \citenamefont
  {Howe}(2018{\natexlab{b}})}]{Agrawal:2017evu}%
  \BibitemOpen
  \bibfield  {author} {\bibinfo {author} {\bibfnamefont {P.}~\bibnamefont
  {Agrawal}}\ and\ \bibinfo {author} {\bibfnamefont {K.}~\bibnamefont {Howe}},\
  }\href {\doibase 10.1007/JHEP12(2018)035} {\bibfield  {journal} {\bibinfo
  {journal} {JHEP}\ }\textbf {\bibinfo {volume} {12}},\ \bibinfo {pages} {035}
  (\bibinfo {year} {2018}{\natexlab{b}})},\ \Eprint
  {http://arxiv.org/abs/1712.05803} {arXiv:1712.05803 [hep-ph]} \BibitemShut
  {NoStop}%
\bibitem [{\citenamefont {Gaillard}\ \emph {et~al.}(2018)\citenamefont
  {Gaillard}, \citenamefont {Gavela}, \citenamefont {Houtz}, \citenamefont
  {Quilez},\ and\ \citenamefont {Del~Rey}}]{Gaillard:2018xgk}%
  \BibitemOpen
  \bibfield  {author} {\bibinfo {author} {\bibfnamefont {M.~K.}\ \bibnamefont
  {Gaillard}}, \bibinfo {author} {\bibfnamefont {M.~B.}\ \bibnamefont
  {Gavela}}, \bibinfo {author} {\bibfnamefont {R.}~\bibnamefont {Houtz}},
  \bibinfo {author} {\bibfnamefont {P.}~\bibnamefont {Quilez}}, \ and\ \bibinfo
  {author} {\bibfnamefont {R.}~\bibnamefont {Del~Rey}},\ }\href {\doibase
  10.1140/epjc/s10052-018-6396-6} {\bibfield  {journal} {\bibinfo  {journal}
  {Eur. Phys. J.}\ }\textbf {\bibinfo {volume} {C78}},\ \bibinfo {pages} {972}
  (\bibinfo {year} {2018})},\ \Eprint {http://arxiv.org/abs/1805.06465}
  {arXiv:1805.06465 [hep-ph]} \BibitemShut {NoStop}%
\bibitem [{\citenamefont {Holdom}\ and\ \citenamefont
  {Peskin}(1982)}]{Holdom:1982ex}%
  \BibitemOpen
  \bibfield  {author} {\bibinfo {author} {\bibfnamefont {B.}~\bibnamefont
  {Holdom}}\ and\ \bibinfo {author} {\bibfnamefont {M.~E.}\ \bibnamefont
  {Peskin}},\ }\href {\doibase 10.1016/0550-3213(82)90228-0} {\bibfield
  {journal} {\bibinfo  {journal} {Nucl. Phys.}\ }\textbf {\bibinfo {volume}
  {B208}},\ \bibinfo {pages} {397} (\bibinfo {year} {1982})}\BibitemShut
  {NoStop}%
\bibitem [{\citenamefont {Flynn}\ and\ \citenamefont
  {Randall}(1987)}]{Flynn:1987rs}%
  \BibitemOpen
  \bibfield  {author} {\bibinfo {author} {\bibfnamefont {J.~M.}\ \bibnamefont
  {Flynn}}\ and\ \bibinfo {author} {\bibfnamefont {L.}~\bibnamefont
  {Randall}},\ }\href {\doibase 10.1016/0550-3213(87)90089-7} {\bibfield
  {journal} {\bibinfo  {journal} {Nucl. Phys.}\ }\textbf {\bibinfo {volume}
  {B293}},\ \bibinfo {pages} {731} (\bibinfo {year} {1987})}\BibitemShut
  {NoStop}%
\bibitem [{\citenamefont {Weinberg}(1978)}]{Weinberg:1977ma}%
  \BibitemOpen
  \bibfield  {author} {\bibinfo {author} {\bibfnamefont {S.}~\bibnamefont
  {Weinberg}},\ }\href {\doibase 10.1103/PhysRevLett.40.223} {\bibfield
  {journal} {\bibinfo  {journal} {Phys. Rev. Lett.}\ }\textbf {\bibinfo
  {volume} {40}},\ \bibinfo {pages} {223} (\bibinfo {year} {1978})}\BibitemShut
  {NoStop}%
\bibitem [{\citenamefont {Wilczek}(1978)}]{Wilczek:1977pj}%
  \BibitemOpen
  \bibfield  {author} {\bibinfo {author} {\bibfnamefont {F.}~\bibnamefont
  {Wilczek}},\ }\href {\doibase 10.1103/PhysRevLett.40.279} {\bibfield
  {journal} {\bibinfo  {journal} {Phys. Rev. Lett.}\ }\textbf {\bibinfo
  {volume} {40}},\ \bibinfo {pages} {279} (\bibinfo {year} {1978})}\BibitemShut
  {NoStop}%
\bibitem [{\citenamefont {Vafa}\ and\ \citenamefont
  {Witten}(1984)}]{Vafa:1984xg}%
  \BibitemOpen
  \bibfield  {author} {\bibinfo {author} {\bibfnamefont {C.}~\bibnamefont
  {Vafa}}\ and\ \bibinfo {author} {\bibfnamefont {E.}~\bibnamefont {Witten}},\
  }\href {\doibase 10.1103/PhysRevLett.53.535} {\bibfield  {journal} {\bibinfo
  {journal} {Phys. Rev. Lett.}\ }\textbf {\bibinfo {volume} {53}},\ \bibinfo
  {pages} {535} (\bibinfo {year} {1984})}\BibitemShut {NoStop}%
\bibitem [{\citenamefont {Callan}\ \emph {et~al.}(1978)\citenamefont {Callan},
  \citenamefont {Dashen},\ and\ \citenamefont {Gross}}]{Callan:1977gz}%
  \BibitemOpen
  \bibfield  {author} {\bibinfo {author} {\bibfnamefont {C.~G.}\ \bibnamefont
  {Callan}, \bibfnamefont {Jr.}}, \bibinfo {author} {\bibfnamefont {R.~F.}\
  \bibnamefont {Dashen}}, \ and\ \bibinfo {author} {\bibfnamefont {D.~J.}\
  \bibnamefont {Gross}},\ }\href {\doibase 10.1103/PhysRevD.17.2717} {\bibfield
   {journal} {\bibinfo  {journal} {Phys. Rev.}\ }\textbf {\bibinfo {volume}
  {D17}},\ \bibinfo {pages} {2717} (\bibinfo {year} {1978})}\BibitemShut
  {NoStop}%
\bibitem [{\citenamefont {'t~Hooft}(1976)}]{tHooft:1976snw}%
  \BibitemOpen
  \bibfield  {author} {\bibinfo {author} {\bibfnamefont {G.}~\bibnamefont
  {'t~Hooft}},\ }\href {\doibase 10.1103/PhysRevD.18.2199.3,
  10.1103/PhysRevD.14.3432} {\bibfield  {journal} {\bibinfo  {journal} {Phys.
  Rev.}\ }\textbf {\bibinfo {volume} {D14}},\ \bibinfo {pages} {3432} (\bibinfo
  {year} {1976})}\BibitemShut {NoStop}%
\bibitem [{\citenamefont {Affleck}(1981)}]{Affleck:1980mp}%
  \BibitemOpen
  \bibfield  {author} {\bibinfo {author} {\bibfnamefont {I.}~\bibnamefont
  {Affleck}},\ }\href {\doibase 10.1016/0550-3213(81)90307-2} {\bibfield
  {journal} {\bibinfo  {journal} {Nucl. Phys.}\ }\textbf {\bibinfo {volume}
  {B191}},\ \bibinfo {pages} {429} (\bibinfo {year} {1981})}\BibitemShut
  {NoStop}%
\bibitem [{\citenamefont {Diakonov}\ and\ \citenamefont
  {Petrov}(1984)}]{Diakonov:1983hh}%
  \BibitemOpen
  \bibfield  {author} {\bibinfo {author} {\bibfnamefont {D.}~\bibnamefont
  {Diakonov}}\ and\ \bibinfo {author} {\bibfnamefont {V.~{\relax Yu}.}\
  \bibnamefont {Petrov}},\ }\href {\doibase 10.1016/0550-3213(84)90432-2}
  {\bibfield  {journal} {\bibinfo  {journal} {Nucl. Phys.}\ }\textbf {\bibinfo
  {volume} {B245}},\ \bibinfo {pages} {259} (\bibinfo {year}
  {1984})}\BibitemShut {NoStop}%
\bibitem [{\citenamefont {Machacek}\ and\ \citenamefont
  {Vaughn}(1983)}]{Machacek:1983tz}%
  \BibitemOpen
  \bibfield  {author} {\bibinfo {author} {\bibfnamefont {M.~E.}\ \bibnamefont
  {Machacek}}\ and\ \bibinfo {author} {\bibfnamefont {M.~T.}\ \bibnamefont
  {Vaughn}},\ }\href {\doibase 10.1016/0550-3213(83)90610-7} {\bibfield
  {journal} {\bibinfo  {journal} {Nucl. Phys.}\ }\textbf {\bibinfo {volume}
  {B222}},\ \bibinfo {pages} {83} (\bibinfo {year} {1983})}\BibitemShut
  {NoStop}%
\bibitem [{\citenamefont {Jones}(1982)}]{Jones:1981we}%
  \BibitemOpen
  \bibfield  {author} {\bibinfo {author} {\bibfnamefont {D.~R.~T.}\
  \bibnamefont {Jones}},\ }\href {\doibase 10.1103/PhysRevD.25.581} {\bibfield
  {journal} {\bibinfo  {journal} {Phys. Rev.}\ }\textbf {\bibinfo {volume}
  {D25}},\ \bibinfo {pages} {581} (\bibinfo {year} {1982})}\BibitemShut
  {NoStop}%
\bibitem [{\citenamefont {Irastorza}\ and\ \citenamefont
  {Redondo}(2018)}]{Irastorza:2018dyq}%
  \BibitemOpen
  \bibfield  {author} {\bibinfo {author} {\bibfnamefont {I.~G.}\ \bibnamefont
  {Irastorza}}\ and\ \bibinfo {author} {\bibfnamefont {J.}~\bibnamefont
  {Redondo}},\ }\href {\doibase 10.1016/j.ppnp.2018.05.003} {\bibfield
  {journal} {\bibinfo  {journal} {Prog. Part. Nucl. Phys.}\ }\textbf {\bibinfo
  {volume} {102}},\ \bibinfo {pages} {89} (\bibinfo {year} {2018})},\ \Eprint
  {http://arxiv.org/abs/1801.08127} {arXiv:1801.08127 [hep-ph]} \BibitemShut
  {NoStop}%
\bibitem [{\citenamefont {Raffelt}(2008)}]{Raffelt:2006cw}%
  \BibitemOpen
  \bibfield  {author} {\bibinfo {author} {\bibfnamefont {G.~G.}\ \bibnamefont
  {Raffelt}},\ }\href {\doibase 10.1007/978-3-540-73518-2_3} {\bibfield
  {journal} {\bibinfo  {journal} {Lect. Notes Phys.}\ }\textbf {\bibinfo
  {volume} {741}},\ \bibinfo {pages} {51} (\bibinfo {year} {2008})},\ \Eprint
  {http://arxiv.org/abs/hep-ph/0611350} {arXiv:hep-ph/0611350 [hep-ph]}
  \BibitemShut {NoStop}%
\bibitem [{\citenamefont {Preskill}\ \emph {et~al.}(1983)\citenamefont
  {Preskill}, \citenamefont {Wise},\ and\ \citenamefont
  {Wilczek}}]{Preskill:1982cy}%
  \BibitemOpen
  \bibfield  {author} {\bibinfo {author} {\bibfnamefont {J.}~\bibnamefont
  {Preskill}}, \bibinfo {author} {\bibfnamefont {M.~B.}\ \bibnamefont {Wise}},
  \ and\ \bibinfo {author} {\bibfnamefont {F.}~\bibnamefont {Wilczek}},\ }\href
  {\doibase 10.1016/0370-2693(83)90637-8} {\bibfield  {journal} {\bibinfo
  {journal} {Phys. Lett.}\ }\textbf {\bibinfo {volume} {120B}},\ \bibinfo
  {pages} {127} (\bibinfo {year} {1983})}\BibitemShut {NoStop}%
\bibitem [{\citenamefont {Abbott}\ and\ \citenamefont
  {Sikivie}(1983)}]{Abbott:1982af}%
  \BibitemOpen
  \bibfield  {author} {\bibinfo {author} {\bibfnamefont {L.~F.}\ \bibnamefont
  {Abbott}}\ and\ \bibinfo {author} {\bibfnamefont {P.}~\bibnamefont
  {Sikivie}},\ }\href {\doibase 10.1016/0370-2693(83)90638-X} {\bibfield
  {journal} {\bibinfo  {journal} {Phys. Lett.}\ }\textbf {\bibinfo {volume}
  {120B}},\ \bibinfo {pages} {133} (\bibinfo {year} {1983})}\BibitemShut
  {NoStop}%
\bibitem [{\citenamefont {Dine}\ and\ \citenamefont
  {Fischler}(1983)}]{Dine:1982ah}%
  \BibitemOpen
  \bibfield  {author} {\bibinfo {author} {\bibfnamefont {M.}~\bibnamefont
  {Dine}}\ and\ \bibinfo {author} {\bibfnamefont {W.}~\bibnamefont
  {Fischler}},\ }\href {\doibase 10.1016/0370-2693(83)90639-1} {\bibfield
  {journal} {\bibinfo  {journal} {Phys. Lett.}\ }\textbf {\bibinfo {volume}
  {120B}},\ \bibinfo {pages} {137} (\bibinfo {year} {1983})}\BibitemShut
  {NoStop}%
\bibitem [{\citenamefont {Wantz}\ and\ \citenamefont
  {Shellard}(2010{\natexlab{a}})}]{Wantz:2009mi}%
  \BibitemOpen
  \bibfield  {author} {\bibinfo {author} {\bibfnamefont {O.}~\bibnamefont
  {Wantz}}\ and\ \bibinfo {author} {\bibfnamefont {E.~P.~S.}\ \bibnamefont
  {Shellard}},\ }\href {\doibase 10.1016/j.nuclphysb.2009.12.005} {\bibfield
  {journal} {\bibinfo  {journal} {Nucl. Phys.}\ }\textbf {\bibinfo {volume}
  {B829}},\ \bibinfo {pages} {110} (\bibinfo {year} {2010}{\natexlab{a}})},\
  \Eprint {http://arxiv.org/abs/0908.0324} {arXiv:0908.0324 [hep-ph]}
  \BibitemShut {NoStop}%
\bibitem [{\citenamefont {Wantz}\ and\ \citenamefont
  {Shellard}(2010{\natexlab{b}})}]{Wantz:2009it}%
  \BibitemOpen
  \bibfield  {author} {\bibinfo {author} {\bibfnamefont {O.}~\bibnamefont
  {Wantz}}\ and\ \bibinfo {author} {\bibfnamefont {E.~P.~S.}\ \bibnamefont
  {Shellard}},\ }\href {\doibase 10.1103/PhysRevD.82.123508} {\bibfield
  {journal} {\bibinfo  {journal} {Phys. Rev.}\ }\textbf {\bibinfo {volume}
  {D82}},\ \bibinfo {pages} {123508} (\bibinfo {year} {2010}{\natexlab{b}})},\
  \Eprint {http://arxiv.org/abs/0910.1066} {arXiv:0910.1066 [astro-ph.CO]}
  \BibitemShut {NoStop}%
\bibitem [{\citenamefont {Gross}\ \emph {et~al.}(1981)\citenamefont {Gross},
  \citenamefont {Pisarski},\ and\ \citenamefont {Yaffe}}]{Gross:1980br}%
  \BibitemOpen
  \bibfield  {author} {\bibinfo {author} {\bibfnamefont {D.~J.}\ \bibnamefont
  {Gross}}, \bibinfo {author} {\bibfnamefont {R.~D.}\ \bibnamefont {Pisarski}},
  \ and\ \bibinfo {author} {\bibfnamefont {L.~G.}\ \bibnamefont {Yaffe}},\
  }\href {\doibase 10.1103/RevModPhys.53.43} {\bibfield  {journal} {\bibinfo
  {journal} {Rev. Mod. Phys.}\ }\textbf {\bibinfo {volume} {53}},\ \bibinfo
  {pages} {43} (\bibinfo {year} {1981})}\BibitemShut {NoStop}%
\bibitem [{\citenamefont {Grilli~di Cortona}\ \emph {et~al.}(2016)\citenamefont
  {Grilli~di Cortona}, \citenamefont {Hardy}, \citenamefont {Pardo~Vega},\ and\
  \citenamefont {Villadoro}}]{diCortona:2015ldu}%
  \BibitemOpen
  \bibfield  {author} {\bibinfo {author} {\bibfnamefont {G.}~\bibnamefont
  {Grilli~di Cortona}}, \bibinfo {author} {\bibfnamefont {E.}~\bibnamefont
  {Hardy}}, \bibinfo {author} {\bibfnamefont {J.}~\bibnamefont {Pardo~Vega}}, \
  and\ \bibinfo {author} {\bibfnamefont {G.}~\bibnamefont {Villadoro}},\ }\href
  {\doibase 10.1007/JHEP01(2016)034} {\bibfield  {journal} {\bibinfo  {journal}
  {JHEP}\ }\textbf {\bibinfo {volume} {01}},\ \bibinfo {pages} {034} (\bibinfo
  {year} {2016})},\ \Eprint {http://arxiv.org/abs/1511.02867} {arXiv:1511.02867
  [hep-ph]} \BibitemShut {NoStop}%
\bibitem [{\citenamefont {Kim}(1979)}]{Kim:1979if}%
  \BibitemOpen
  \bibfield  {author} {\bibinfo {author} {\bibfnamefont {J.~E.}\ \bibnamefont
  {Kim}},\ }\href {\doibase 10.1103/PhysRevLett.43.103} {\bibfield  {journal}
  {\bibinfo  {journal} {Phys. Rev. Lett.}\ }\textbf {\bibinfo {volume} {43}},\
  \bibinfo {pages} {103} (\bibinfo {year} {1979})}\BibitemShut {NoStop}%
\bibitem [{\citenamefont {Shifman}\ \emph {et~al.}(1980)\citenamefont
  {Shifman}, \citenamefont {Vainshtein},\ and\ \citenamefont
  {Zakharov}}]{Shifman:1979if}%
  \BibitemOpen
  \bibfield  {author} {\bibinfo {author} {\bibfnamefont {M.~A.}\ \bibnamefont
  {Shifman}}, \bibinfo {author} {\bibfnamefont {A.~I.}\ \bibnamefont
  {Vainshtein}}, \ and\ \bibinfo {author} {\bibfnamefont {V.~I.}\ \bibnamefont
  {Zakharov}},\ }\href {\doibase 10.1016/0550-3213(80)90209-6} {\bibfield
  {journal} {\bibinfo  {journal} {Nucl. Phys.}\ }\textbf {\bibinfo {volume}
  {B166}},\ \bibinfo {pages} {493} (\bibinfo {year} {1980})}\BibitemShut
  {NoStop}%
\bibitem [{\citenamefont {Zhitnitsky}(1980)}]{Zhitnitsky:1980tq}%
  \BibitemOpen
  \bibfield  {author} {\bibinfo {author} {\bibfnamefont {A.~R.}\ \bibnamefont
  {Zhitnitsky}},\ }\href@noop {} {\bibfield  {journal} {\bibinfo  {journal}
  {Sov. J. Nucl. Phys.}\ }\textbf {\bibinfo {volume} {31}},\ \bibinfo {pages}
  {260} (\bibinfo {year} {1980})},\ \bibinfo {note} {[Yad.
  Fiz.31,497(1980)]}\BibitemShut {NoStop}%
\bibitem [{\citenamefont {Chikashige}\ \emph {et~al.}(1981)\citenamefont
  {Chikashige}, \citenamefont {Mohapatra},\ and\ \citenamefont
  {Peccei}}]{Chikashige:1980ui}%
  \BibitemOpen
  \bibfield  {author} {\bibinfo {author} {\bibfnamefont {Y.}~\bibnamefont
  {Chikashige}}, \bibinfo {author} {\bibfnamefont {R.~N.}\ \bibnamefont
  {Mohapatra}}, \ and\ \bibinfo {author} {\bibfnamefont {R.~D.}\ \bibnamefont
  {Peccei}},\ }\href {\doibase 10.1016/0370-2693(81)90011-3} {\bibfield
  {journal} {\bibinfo  {journal} {Phys. Lett.}\ }\textbf {\bibinfo {volume}
  {98B}},\ \bibinfo {pages} {265} (\bibinfo {year} {1981})}\BibitemShut
  {NoStop}%
\bibitem [{\citenamefont {Gelmini}\ and\ \citenamefont
  {Roncadelli}(1981)}]{Gelmini:1980re}%
  \BibitemOpen
  \bibfield  {author} {\bibinfo {author} {\bibfnamefont {G.~B.}\ \bibnamefont
  {Gelmini}}\ and\ \bibinfo {author} {\bibfnamefont {M.}~\bibnamefont
  {Roncadelli}},\ }\href {\doibase 10.1016/0370-2693(81)90559-1} {\bibfield
  {journal} {\bibinfo  {journal} {Phys. Lett.}\ }\textbf {\bibinfo {volume}
  {99B}},\ \bibinfo {pages} {411} (\bibinfo {year} {1981})}\BibitemShut
  {NoStop}%
\bibitem [{\citenamefont {Kim}(1981)}]{Kim:1981jw}%
  \BibitemOpen
  \bibfield  {author} {\bibinfo {author} {\bibfnamefont {J.~E.}\ \bibnamefont
  {Kim}},\ }\href {\doibase 10.1016/0370-2693(81)91149-7} {\bibfield  {journal}
  {\bibinfo  {journal} {Phys. Lett.}\ }\textbf {\bibinfo {volume} {107B}},\
  \bibinfo {pages} {69} (\bibinfo {year} {1981})}\BibitemShut {NoStop}%
\bibitem [{\citenamefont {Langacker}\ \emph {et~al.}(1986)\citenamefont
  {Langacker}, \citenamefont {Peccei},\ and\ \citenamefont
  {Yanagida}}]{Langacker:1986rj}%
  \BibitemOpen
  \bibfield  {author} {\bibinfo {author} {\bibfnamefont {P.}~\bibnamefont
  {Langacker}}, \bibinfo {author} {\bibfnamefont {R.~D.}\ \bibnamefont
  {Peccei}}, \ and\ \bibinfo {author} {\bibfnamefont {T.}~\bibnamefont
  {Yanagida}},\ }\href {\doibase 10.1142/S0217732386000683} {\bibfield
  {journal} {\bibinfo  {journal} {Mod. Phys. Lett.}\ }\textbf {\bibinfo
  {volume} {A1}},\ \bibinfo {pages} {541} (\bibinfo {year} {1986})}\BibitemShut
  {NoStop}%
\bibitem [{\citenamefont {Ma}(2006)}]{Ma:2006km}%
  \BibitemOpen
  \bibfield  {author} {\bibinfo {author} {\bibfnamefont {E.}~\bibnamefont
  {Ma}},\ }\href {\doibase 10.1103/PhysRevD.73.077301} {\bibfield  {journal}
  {\bibinfo  {journal} {Phys. Rev.}\ }\textbf {\bibinfo {volume} {D73}},\
  \bibinfo {pages} {077301} (\bibinfo {year} {2006})},\ \Eprint
  {http://arxiv.org/abs/hep-ph/0601225} {arXiv:hep-ph/0601225 [hep-ph]}
  \BibitemShut {NoStop}%
\bibitem [{\citenamefont {Ma}\ \emph {et~al.}(2017)\citenamefont {Ma},
  \citenamefont {Ohata},\ and\ \citenamefont {Tsumura}}]{Ma:2017vdv}%
  \BibitemOpen
  \bibfield  {author} {\bibinfo {author} {\bibfnamefont {E.}~\bibnamefont
  {Ma}}, \bibinfo {author} {\bibfnamefont {T.}~\bibnamefont {Ohata}}, \ and\
  \bibinfo {author} {\bibfnamefont {K.}~\bibnamefont {Tsumura}},\ }\href
  {\doibase 10.1103/PhysRevD.96.075039} {\bibfield  {journal} {\bibinfo
  {journal} {Phys. Rev.}\ }\textbf {\bibinfo {volume} {D96}},\ \bibinfo {pages}
  {075039} (\bibinfo {year} {2017})},\ \Eprint
  {http://arxiv.org/abs/1708.03076} {arXiv:1708.03076 [hep-ph]} \BibitemShut
  {NoStop}%
\bibitem [{\citenamefont {Reig}\ \emph {et~al.}(2018)\citenamefont {Reig},
  \citenamefont {Restrepo}, \citenamefont {Valle},\ and\ \citenamefont
  {Zapata}}]{Reig:2018mdk}%
  \BibitemOpen
  \bibfield  {author} {\bibinfo {author} {\bibfnamefont {M.}~\bibnamefont
  {Reig}}, \bibinfo {author} {\bibfnamefont {D.}~\bibnamefont {Restrepo}},
  \bibinfo {author} {\bibfnamefont {J.~W.~F.}\ \bibnamefont {Valle}}, \ and\
  \bibinfo {author} {\bibfnamefont {O.}~\bibnamefont {Zapata}},\ }\href
  {\doibase 10.1103/PhysRevD.97.115032} {\bibfield  {journal} {\bibinfo
  {journal} {Phys. Rev.}\ }\textbf {\bibinfo {volume} {D97}},\ \bibinfo {pages}
  {115032} (\bibinfo {year} {2018})},\ \Eprint
  {http://arxiv.org/abs/1803.08528} {arXiv:1803.08528 [hep-ph]} \BibitemShut
  {NoStop}%
\bibitem [{\citenamefont {Hehn}\ and\ \citenamefont
  {Ibarra}(2013)}]{Hehn:2012kz}%
  \BibitemOpen
  \bibfield  {author} {\bibinfo {author} {\bibfnamefont {D.}~\bibnamefont
  {Hehn}}\ and\ \bibinfo {author} {\bibfnamefont {A.}~\bibnamefont {Ibarra}},\
  }\href {\doibase 10.1016/j.physletb.2012.11.034} {\bibfield  {journal}
  {\bibinfo  {journal} {Phys. Lett.}\ }\textbf {\bibinfo {volume} {B718}},\
  \bibinfo {pages} {988} (\bibinfo {year} {2013})},\ \Eprint
  {http://arxiv.org/abs/1208.3162} {arXiv:1208.3162 [hep-ph]} \BibitemShut
  {NoStop}%
\end{thebibliography}%

\end{document}